\documentclass[%
 reprint,floatfix,
 amsmath,amssymb,
 aps,
]{revtex4-2}

\usepackage{graphicx}% Include figure files
\usepackage{dcolumn}% Align table columns on decimal point
\usepackage{bm}% bold math
\usepackage{braket}
\usepackage{orcidlink,mathrsfs,here}
\usepackage[dvipsnames]{xcolor}
\usepackage[compat=1.1.0]{tikz-feynhand}
\usetikzlibrary{positioning}
\usetikzlibrary{intersections,calc,arrows.meta}

\usepackage{CJKutf8}

\usepackage{hyperref}% add hypertext capabilities
\hypersetup{%
 setpagesize=false,%
 bookmarks=true,%
 bookmarksdepth=tocdepth,%
 bookmarksnumbered=true,%
 colorlinks=false,%
 pdftitle={},%
 pdfsubject={},%
 pdfauthor={},%
 pdfkeywords={}}
\newcommand{\Tcc}{T_{cc}(3875)^+}
\newcommand{\ccud}{cc\bar{u}\bar{d}}

\newcommand{\uc}{D^{*0}D^+}
\newcommand{\lc}{D^{*+}D^0}

\newcommand{\DDp}{D^0D^0\pi^+}

\newcommand{\DDg}{D^0D^+\gamma}
  
\begin{document}

\preprint{APS/123-QED}
\title{A Study of $\Tcc$ Nature : Compact v.s. Molecule}


\author{Shota Ampuku\,\orcidlink{0009-0005-3588-7806}}
\email{ampuku@hken.phys.nagoya-u.ac.jp}
\affiliation{Department of Physics, Nagoya University, Nagoya 464-8602, Japan}
\author{Yasuhiro Yamaguchi\,\orcidlink{0000-0003-1347-0821}}
\email{yamaguchi@hken.phys.nagoya-u.ac.jp}
\affiliation{Department of Physics, Nagoya University, Nagoya 464-8602, Japan}
\affiliation{Kobayashi-Maskawa Institute for the Origin of Particles and the Universe, Nagoya University, Nagoya, 464-8602, Japan}
\author{Masayasu Harada}
\email{harada@hken.phys.nagoya-u.ac.jp}
\affiliation{Kobayashi-Maskawa Institute for the Origin of Particles and the Universe, Nagoya University, Nagoya, 464-8602, Japan}
\affiliation{Department of Physics, Nagoya University, Nagoya 464-8602, Japan}
\affiliation{Advanced Science Research Center, Japan Atomic Energy Agency, Tokai 319-1195, Japan}

\date{\today}% It is always \today, today,

\begin{abstract}
A central question in exotic-hadron physics is their internal structure whether these states are loosely bound hadronic molecules or compact multiquark configurations.
To shed light on this issue, we develop a model that incorporates mixing between hadronic-molecular and compact multiquark components.
We then apply this framework to the specific case of the $\Tcc$ and analyze the peak structure in the $\DDp$ invariant-mass spectrum reported by LHCb.
We find that a scenario based on a predominantly compact tetraquark provides the best fitted solution which can explain the $\Tcc$. 
We also find that the model admits two more solutions of comparable quality, both of which imply that the $\Tcc$ is a molecular state: (1) the $\Tcc$ is a $\lc$ molecule and there is a $\uc$ molecular state in addition; (2) the $\Tcc$ is a $\uc$ molecule and an aditional $\lc$ molecular state is found below $\DDp$ threshold.
These molecular states are not simple $I = 0$ states, but mixtures of $I = 0$ and $I = 1$ states.
We show that all three scenarios are also consistent with the experimentally observed near-threshold $D^0D^0$ and $D^0D^+$ invariant-mass distributions.

\end{abstract}

\maketitle

\textit{Introduction}---
Hadrons are composite particles bound by the strong interaction between quarks and gluons, traditionally classified as mesons (quark-antiquark states) or baryons (three-quark states). 
Beyond this meson-baryon picture, a class of states collectively referred to as “exotic hadrons,” whose properties are difficult to accommodate within the conventional picture, has become a central focus of contemporary hadron physics. 
Since the 2000s, numerous candidates have been reported experimentally,
 and intensively analyzed by both experimental and theoretical communities. 
(For recent progress, see, e.g., the reviews \cite{Chen:2022asf,Liu:2024uxn} and references therein.)

Despite the wealth of candidates, the fundamental question—what microscopic configuration these states realize—remains open.
Interpretations of the internal structure span loosely bound hadronic molecules, compact multiquarks, hybrids, glueballs, and even non-resonant kinematic effects.
Determining the internal structure would reveal nontrivial color configurations not seen in ordinary hadrons and, by directly accessing non-perturbative QCD, advance our understanding of its dynamics in the low energy region.

A notable example is the narrow peak reported by the LHCb Collaboration in 2021 in the $\DDp$ invariant mass distribution in pp collisions \cite{LHCb2021vvq,LHCb:2021auc}, commonly identified with the doubly charmed tetraquark $\Tcc$.
The mass difference from the $\lc$ threshold and decay width as an amplitude-pole are $\delta m_{\rm{pole}}=-359\pm 40^{+4}_{-0}\,\rm{keV/c^2}$, $\Gamma_{\rm{pole}}=-48\pm 2^{+0}_{-14}\,\rm{keV}$.
To analyze the data, a unitarized Breit-Wigner model—preserving unitarity and analyticity—was fitted to the $\DDp$ spectrum, based on the following assumptions:
(i) The observed state is the $\ccud$ ground state with quantum numbers $I(J^P)=0(1^+)$.
(ii) Owing to its proximity to the $DD^*$ threshold, $\Tcc$ strongly couples to the $DD^*$ channel.
These assumptions are supported by the simultaneous description of $\DDp$ and $D^0\pi^+$ mass distributions, as well as the $D^0D^0$ and $D^0D^+$ invariant mass spectra.

The peak structure in the $\DDp$ line shape—regarded as the $\ccud$ ground state—lies just below the $DD^*$ thresholds. 
This observation indicates that a hadronic molecular model naturally accounts for a near-threshold $\Tcc$ as a shallow $DD^*$ bound state.
(See, e.g., Refs.
\cite{Sakai:2025djx,Chen:2022asf} and references therein.)
Nevertheless, the possibility that $\ccud$ forms a compact tetraquark has not been ruled out.
In fact, models with explicit quark degrees of freedom can reproduce the LHCb mass either by tuning the model parameters within the theoretical uncertainties or by introducing additional interaction terms.
(See, e.g., Refs. \cite{He:2023ucd,Ma:2023int,Meng:2023jqk}.)

These studies have been performed independently.
However, it should be emphasized that a physical state can emerge as a quantum-mechanical admixture of a compact-multiquark and a hadronic-molecular components \cite{Weinberg:1962hj,Kinugawa:2024crb}.
This situation finds a close parallel in the case of the $X(3872)$, where a non-negligible compact core appears to coexist with a dominant $D\bar D^*$ molecular component.
The inclusion of the compact core has been suggested by production and radiative decay data which are difficult to be described within a purely molecular model. 
(See, e.g., Refs.\cite{Yamaguchi:2019vea,Esposito:2025hlp}.)
In general, such compact-molecular admixtures are expected for multiquark candidates.

As discussed in the case of $X(3872)$, the observed mass alone is not sufficient to determine the $T_{cc}^+$ spatial configuration; decisive tests would come from the near-threshold line shapes—whose slopes, shoulders, and interference patterns include the information about the internal structure.
While the $\DDp$ line shape has predominantly been analyzed under a purely molecular assumption \cite{Albaladejo:2021vln,Du:2021zzh,Qiu:2023uno,Wang:2023ovj,Zhang:2024dth}, the studies do not consider mixing with a compact tetraquark component.
These considerations call for an analysis of the near-threshold line shapes in a framework that permits molecular-compact mixing, allowing the compact effects to be taken into consideration by data rather than by assumption.

In this study, we develop a model that describes the $T_{cc}^+$ as a quantum superposition of hadronic-molecular and compact-tetraquark components.
We fit the model parameters to the 200\,keV-binned $\DDp$ line shape data \cite{LHCb2021vvq,LHCb:2021auc} and analyze the $T_{cc}^+$ properties.
In addition, by using the obtained parameters, we predict the $D^0D^0$ and $D^0D^+$ invariant-mass distributions and compare the LHCb data to verify the results.

\textit{Basic Framework}---
We develop a low-energy effective model for $DD^*$ interactions in the near-threshold region, where the dynamics are governed predominantly by s-wave scattering.
To capture the salient properties of the $T_{cc}^+$ resonance emerging close to the $DD^*$ thresholds, we adopt a contact-interaction framework and introduce an explicit compact tetraquark field, $T_{cc}^+$, which couples to the $DD^*$ continuum states.
Within the approach of Ref. \cite{Yamaguchi:2016kxa}, the interaction Lagrangian involving $DD^*$ pairs and the compact $T_{cc}^+$ field is written as
\vspace{-3pt}
\begin{equation}
  \begin{aligned}
    \mathcal{L}_{\textrm{int}}
        =&
        -\sum_{I=0,1} C_I \left[DD^*\right]^\dagger_I \left[DD^*\right]_I\\
        &-g_0
        \left(
          [DD^*]^\dagger_{I=0}T_{cc}+T^\dagger_{cc} [DD^*]_{I=0}  
        \right),
  \end{aligned}
\end{equation}
where the coupling constants $C_I$ parametrize the low-energy interactions in the isospin $I=0,1$ channels, and $g_0$ denotes the coupling between the compact $T_{cc}^+$ and the isoscalar $DD^*$ state.

The states of $\lc$ and $\uc$ in the particle basis are built from the combinations of the isoscalar and isovector $DD^*$ as
\begin{eqnarray}
  \ket{\lc}&=-\dfrac{1}{\sqrt{2}}\left(\ket{DD^*,I=0}+\ket{DD^*,I=1}\right),\\
  \ket{\uc}&=-\dfrac{1}{\sqrt{2}}\left(\ket{DD^*,I=0}-\ket{DD^*,I=1}\right).
\end{eqnarray}
In the particle basis, the resulting interaction potential (after an appropriate rescaling) takes the form
\begin{equation}
    \begin{aligned}
    V_{ij} =& \frac{1}{2}
    \begin{pmatrix}
    C_1 + C_0 & C_1 - C_0 \\ C_1 - C_0 & C_1 + C_0
    \end{pmatrix}\\
    &+ \frac{1}{2} \frac{1}{E - m_{4q}}
    \begin{pmatrix}
    g_0^2 & -g_0^2 \\ -g_0^2 & g_0^2
    \end{pmatrix},\label{eq:pot}
    \end{aligned}
\end{equation}
where $m_{4q}$ represents the mass of the compact $T_{cc}^+$ state.
Here, the channel indices are defined such that $i = 1$ corresponds to the $\lc$ channel and $i = 2$ corresponds to the $\uc$ channel.
To analyze the near-threshold s-wave coupled-channel dynamics, we consider the coupled-channel Lippmann-Schwinger equation in the center-of-mass frame of a two-meson system with the total energy $E$:
\begin{equation}
  T_{ij}(E)=V_{ij}(E)+V_{ik}(E)G_{k}(E)T_{kj}(E),
\end{equation}
where repeated indices imply summation over intermediate channels.
The two-body propagator $G_i(E)$ for channel $i$ is defined as
\begin{equation}
  \begin{aligned}
      G_i(E)
      =\int_{0}^{\Lambda}\dfrac{d^3k}{(2\pi)^3}\dfrac{1}{E-E_\textrm{th,i}-k^2/2\mu_i+i\varepsilon}.
  \end{aligned}
\end{equation}
Here, $E_{\textrm{th},i}$ denotes the threshold energy of channel $i$, $\mu_i$ is the reduced mass of the two particles in channel $i$, $k$ is the relative momentum between the two particles in the center-of-mass frame, and $\varepsilon$ is an infinitesimal positive number that specifies the causal prescription of the propagator.
This integral exhibits an ultraviolet divergence at high momenta; therefore, we introduce a sharp cutoff $\Lambda$ to tame the divergence.
To account for the finite decay width of the $D^*$, the $D^*$ mass in the loop function is replaced as $m_{D^*}\to m_{D^*}-i\Gamma_{D^*}/2$.

To describe the $\DDp$ line shape reported by LHCb, we consider the following $\DDp$ amplitude \cite{Albaladejo:2021vln}:
\begin{widetext}  
  \begin{equation}
      \mathcal{M}_f(s,t,u;Q^2,\lambda)
      =g_{D^*D\pi}p^\nu_\pi\varepsilon^\mu_s(\lambda)
      \left[\frac{K_t(Q^2)}{t-m^2_{D^*_{(t)}}}
      \left(-g_{\mu\nu}+\frac{k^{(t)}_\mu k^{(t)}_\nu}{m^2_{D^*_{(t)}}}\right)
      +\frac{K_u(Q^2)}{u-m^2_{D^*_{(u)}}}
      \left(-g_{\mu\nu}+\frac{k^{(u)}_\mu k^{(u)}_\nu}{m^2_{D^*_{(u)}}}\right)
      \right],
  \end{equation}
\end{widetext}
where $g_{D^*D\pi}$ is the $D^*\to D\pi$ coupling constant, $p_\pi^\nu$ is the pion four-momentum, $\varepsilon^\mu_s(\lambda)$ is the polarization vector of the $DD^*$ source.
Here, the total four-momentum supplied by the source is $Q^\mu \equiv p_1^\mu+p_2^\mu+p_\pi^\mu$, so that $Q^2=(p_1+p_2+p_\pi)^2$ is the squared invariant mass of the $DD\pi$ system. 
$p_1^\mu$ and $p_2^\mu$ denote the four-momenta of the two final-state $D$ mesons.
For convenience in the $t$- and $u$-channel terms we further introduce $k^{(t)\mu}\equiv p_1^\mu+p_\pi^\mu$ and $k^{(u)\mu}\equiv p_2^\mu+p_\pi^\mu$.
The invariant mass squared $s,t$ and $u$ are defined as
$ s=(p_1+p_2)^2,\,
  t=\big(k^{(t)}\big)^2,\,
  u=\big(k^{(u)}\big)^2$
and they satisfy the relation: $s+t+u=Q^2+p^2_1+p^2_2+p^2_\pi$.
The $K_{t,u}$ functions are introduced to account for the $\lc-\uc$ dynamics.
In the $\DDp$ case, $K_{t,u}$ are symmetric under the exchange $D^0(p_1)\leftrightarrow D^0(p_2)$.
Hence, $K_t=K_u=K_{t,u}$, which can be written as 
\begin{eqnarray}
  K_{t,u}(Q^2)
  =
  \alpha\left(1+G_1(Q^2)T_{11}(Q^2)\right)\notag\\
  +\beta G_2(Q^2)T_{12}(Q^2).
\end{eqnarray}
Here, $\alpha$ and $\beta$ denote the production rates in the respective channels, incorporating the effects of isospin-symmetry breaking near the threshold.
This parametrization includes both the tree-level production mechanism, which serves as a non-resonant background in the $\DDp$ spectrum, and the $DD^*$ rescattering through the loop terms where the resonance is dynamically generated.
From the above, $\DDp$ invariant mass distribution is written as 
\begin{widetext}
  \begin{equation}
      \begin{aligned}
        \mathcal{N}_\text{ev}(Q^2)
        &=\mathcal{N}_0\left(\frac{Q^2_\text{th}}{Q^2}\right)^{\frac{3}{2}}
        \int_{s_\text{th}}^{s_\text{max}(Q^2)}ds
        \int_{t_-\left(s,Q^2\right)}^{t_+\left(s,Q^2\right)}dt\sum_{\lambda}\left|\mathcal{M}_f(s,t,u;Q^2,\lambda)\right|^2.
      \end{aligned}
  \end{equation}
\end{widetext}
Here, the integration range is determined by the Dalitz boundaries.
Furthermore, to account for the experimental resolution, we convolute our theoretical invariant-mass distribution with the resolution function \cite{LHCb:2021auc}.
$D^0D^0$ and $D^0D^+$ distributions are obtained as
\begin{widetext}
  \begin{equation}
    \begin{aligned}
      R_{f}(m_{DD})=
      2m_{DD}
      \int_{(m_\pi+m_{DD})}^{\infty}dQ f_c(Q)\frac{1}{Q}
      \int_{t_-(s,Q^2)}^{t_+(s,Q^2)}dt
      \left|\mathcal{M}_f(s,t,u;Q^2,\lambda)\right|^2
    \end{aligned}
  \end{equation}
\end{widetext}
where $f_c(Q)$ is a Gaussian cutoff function, given in the LHCb analysis procedure \cite{LHCb:2021auc}.

\textit{Results and Discussions}---
We perform a $\chi^2$ fit to the LHCb $\DDp$ line-shape data.
In this analysis, we restrict the fit to the energy region $3.873 < m_{D^0 D^0 \pi^+} < 3.877$ GeV, using the dataset binned at 200 keV.
This choice reflects the limited domain of validity of our effective model, which is formulated for sufficiently low relative momenta.
The complete 500 keV-binned spectrum is not used in our fit, as our effective description is not reliable over that range.

Throughout the input masses and widths, summarized in Tabel \ref{tbl:phys_inputs}, are adopted from the PDG \cite{ParticleDataGroup:2024cfk} except for the $\Gamma_{D^{*0}}$ which is obtained from $\Gamma_{D^{*+}}$ by the isospin symmetry.
Global factors such as the $g_{D^{*}D\pi}$ coupling are absorbed into the normalization constant $\mathcal N_0$ and are not fitted independently.

By fitting to this restricted region, we obtain several parameter sets that reproduce the experimental spectrum with comparable quality.
We present the following three scenarios:
(i) Mol.+Compact (M+C), which includes both hadronic molecular and compact tetraquark components;
(ii) Mol.1, in which a $\lc$ molecule predominantly accounts for the $\DDp$ peak structure; and
(iii) Mol.2, where a $\uc$ molecule does so.
For each scenario, we report the best-fitted parameters obtained by $\chi^2$ minimization in Table \ref{tbl:Best Fitted parameters}.

The parameter set comprises the four-point contact couplings $C_0$ and $C_1$, the compact $T_{cc}^+$ mass $m_{4q}$, the coupling $g_0$ between the compact $T_{cc}^+$ and the $I=0\,DD^*$ scattering states, the production ratio $\beta/\alpha$, and an overall normalization constant $\mathcal N_0$.
\onecolumngrid
    \begin{center}
        \begin{table}[bth]
          \renewcommand{\arraystretch}{1.15}
          \setlength{\tabcolsep}{6pt}
          \begin{tabular}{cccccccc}
            \hline\hline
            $m_{D^0}$ & $m_{D^+}$ & $m_{D^{*0}}$ & $m_{D^{*+}}$ & $m_{\pi^+}$ & $m_{\pi^0}$ & $\Gamma_{D^{*+}}$ & $\Gamma_{D^{*0}}$ \\
            \hline
            $1864.84\,\mathrm{MeV}$ & $1869.66\,\mathrm{MeV}$ & $2006.85\,\mathrm{MeV}$ & $2010.26\,\mathrm{MeV}$ & $139.570\,\mathrm{MeV}$ & $134.977\,\mathrm{MeV}$ & $83.4\,\mathrm{keV}$ & $56.2\,\mathrm{keV}$ \\
            \hline\hline
          \end{tabular}
          \caption{Physical inputs used in the analysis. Masses are PDG world averages; $\Gamma_{D^{*+}}$ is the PDG value; $\Gamma_{D^{*0}}$ is obtained from $\Gamma_{D^{*+}}$ using the isospin symmetry.}
          \label{tbl:phys_inputs}
        \end{table}
    \end{center}
\twocolumngrid
\onecolumngrid
\begin{center}
    \begin{table}[bth]
      \begin{tabular}{c|cccccc}
      Model &$ \beta/\alpha$ & $C_0\,[\textrm{GeV}^{-2}]$ & $C_1\,[\textrm{GeV}^{-2}]$ & $g_0\,[\textrm{GeV}^{-1/2}]$ & $m_{4q}\,[\textrm{GeV}]$ & $\chi^2/\textrm{d.o.f.}$ \\
      \hline
      $\textrm{Mol.+Compact (M+C)}$ & $0.29$ & $ -1.38 $ & $ -0.31 $ & $\pm 0.14$ & $3.8757$ & $15.0/(19-6)=1.15$ \\
      $\textrm{Mol.\,1}$ & $0.058$ & $-22.9$ & $-20.2$ & - & - & $16.0/(19-6)=1.23$ \\
      $\textrm{Mol.\,2}$ & $0.93$ & $ -29.5 $ & $ -23.6 $ & - & - & $15.9/(19-6)=1.22$ \\
      \end{tabular}
      \caption{Best-fitted parameters for the three scenarios.
    $\beta/\alpha$ is the relative production weight of the upper to lower channel.
    $C_{0,1}$ are low-energy contact coupling constants.
    The compact-state coupling $g_0$ and the compact mass $m_{4q}$ appear only in the Mol.+Compact (M+C) scenario; in the pure molecular scenarios (Mol.1, Mol.2), the compact $T_{cc}^+$ state is effectively decoupled, then we do not show these values.
    The rightmost column shows $\chi^2$ per degree of freedom, computed as $\chi^2/(\textrm{d.o.f.})$ with $19$ data points and $6$ fit parameters.
    }
      \label{tbl:Best Fitted parameters}
\end{table}
\end{center}
\twocolumngrid

\begin{figure}[htb]
    \centering
    \includegraphics[width=0.9\linewidth]{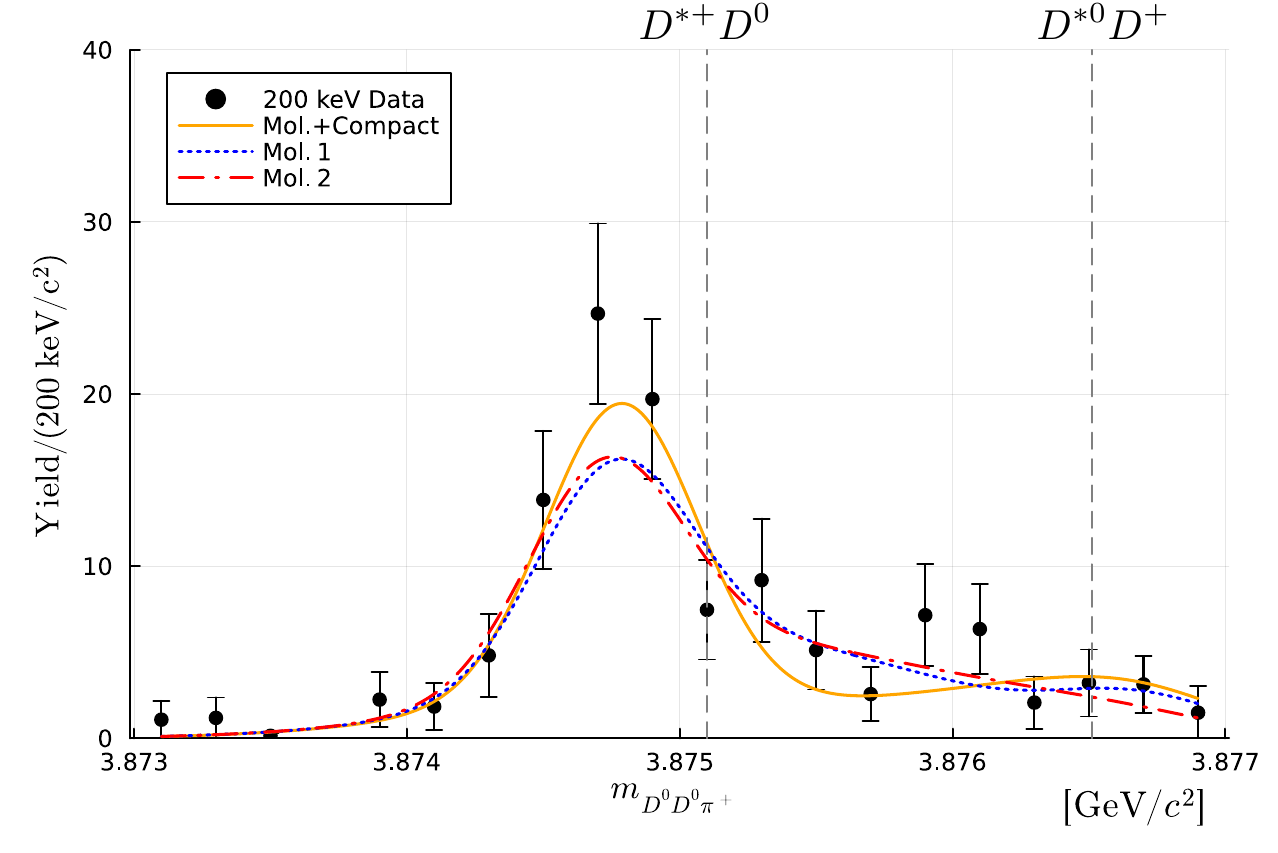}
    \caption{Fitted results of $\DDp$ invariant mass spectrum of each three scenarios:
    Mol.+Compact (solid, orange), Mol.\,1 (dotted, blue), and Mol.\,2 (dashed, red) with the parameters in Table\,\ref{tbl:Best Fitted parameters}.
    Experimental 200\,keV binned $\DDp$ data are shown by black points with statistical error bars.
    Vertical gray dashed lines mark the $\lc$ and $\uc$ thresholds.
    }
    \label{fig:lineshape}
\end{figure}

All parameters are determined by minimizing $\chi^2/\mathrm{d.o.f.}$ against the experimental spectrum; the resulting best-fit $\DDp$ line shapes are shown in Fig.\,\ref{fig:lineshape}.
All three scenarios exhibit a resonance-like enhancement in the $\DDp$ line shape. 
However, mass spectra of the tetraquark that are origins of the enhancement are entirely different one another as follows.

\begin{figure}[htb]
    \centering
    \includegraphics[width=1.0\linewidth]{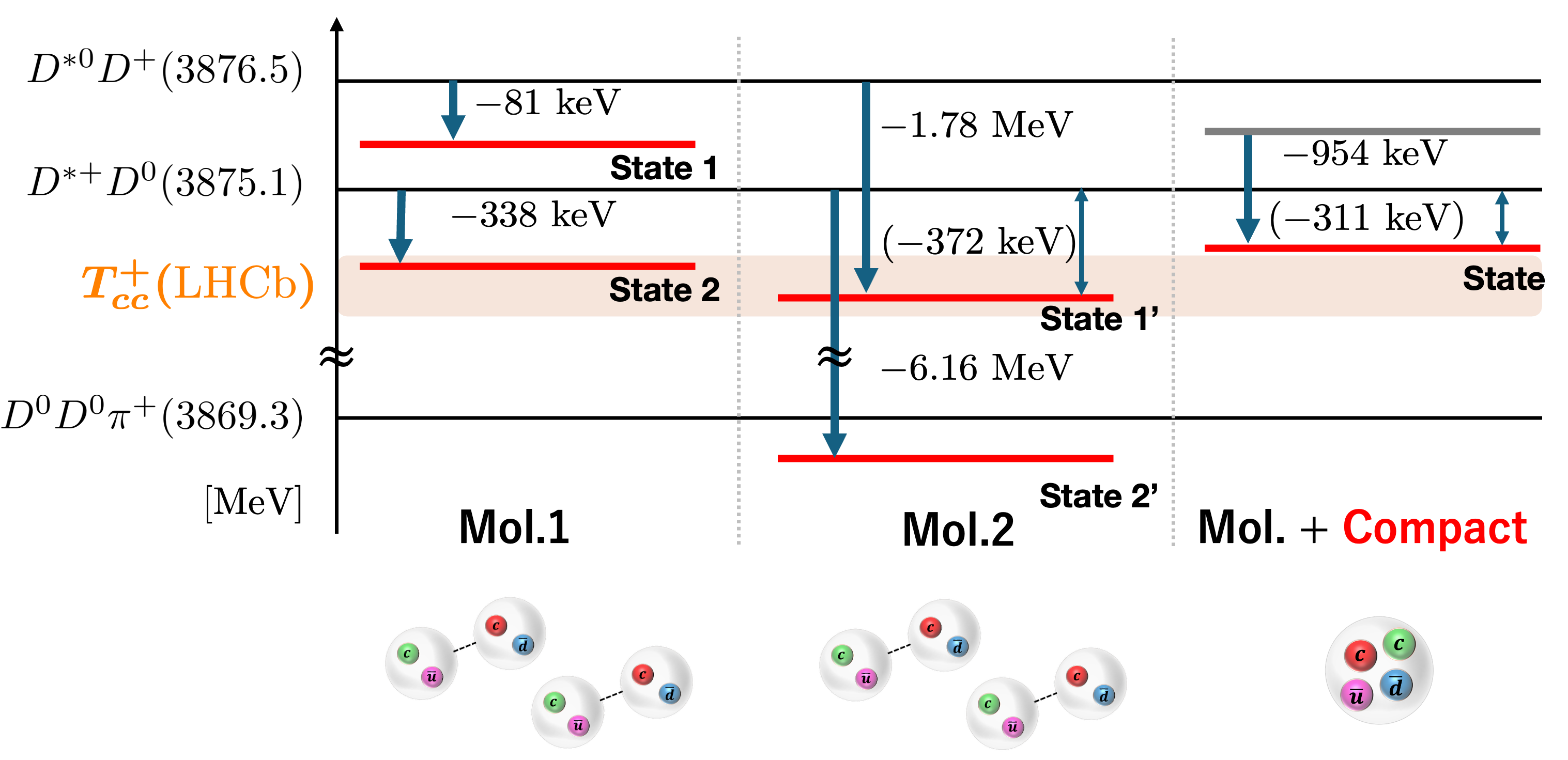}
    \caption{The binding energies obtained in each scenario.
    Red horizontal bars indicate the pole locations.
    Two poles are obtained in Mol.\,1 and Mol.\,2, while only one single near-threshold pole is found in the Mol.+Compact scenario.
    Blue arrows annotate the binding energies measured from the threshold of the dominant channel assigned analyzing the residue of the scattering amplitude.
    For the M+C scenario, the value $-954\,$ keV is measured from the $m_{4q}$.
    The values shown in parentheses are the differences from the $\lc$ threshold.
    The orange band shows the LHCb determination of mass of $T_{cc}^+$.
    }
    \label{fig:spectrum}
\end{figure}

\begin{enumerate}
\item[M+C:] As shown in Fig.\,\ref{fig:lineshape} (orange solid curve), the $\DDp$ line shape exhibits a resonance-like enhancement. 
    In fact, we find a bound state just 311\,keV below the $\lc$ threshold, as shown in Fig.\,\ref{fig:spectrum}. 
    The resonance-like structure in Fig.\,\ref{fig:lineshape} is mainly driven by the compact $T_{cc}^+$ state located above the $\lc$ threshold, because the structure disappears when the compact state is decoupled in the limit of $g_0 \to 0$. 
    The small strengths of $C_0$ and $C_1$ also indicate that the $DD^*$ interaction is weak and not enough to make a $DD^*$ molecular state. However, the coupling to the $DD^*$ channel shifts the compact state  downward by 954\,keV, and the physical $T_{cc}^+$ state is obtained below the $\lc$ threshold.
\end{enumerate}
    We obtained a best fit in the Mol.\,+\,Compact model that prefers a predominantly compact configuration. 
    On the other hand, we also find two pure-molecular scenarios whose natures are completely different from the Mol.\,+\,Compact one.
    Fit qualities of these two scenarios, Mol.\,1 and Mol.\,2, are comparable to that of the Mol.\,+\,Compact scenario.
    In these scenarios, the compact $T_{cc}^+$ is effectively decoupled from the $DD^*$ state, and thus is negligible for the dynamics.
\begin{enumerate}
    \item[Mol.\,1:] 
    In this scenario, we obtain two bound states, State 1 and State 2, as shown in Fig. \ref{fig:spectrum}. 
    State 1 is located 81 keV below the $\uc$ threshold, and State 2 is located 338 keV below the $\lc$ one corresponding to the experimentally found peak in the $\DDp$ line shape.
    By analyzing pole resides of the two-body scattering amplitude, we find that the State 1 and State 2 are realized as the $\uc$ and $\lc$ bound states, respectively.
    This implies that the $\Tcc$ is not a simple $I = 0$ state, but a mixture of $I = 0$ and $I=1$ state.
    The reason why a peak of State 1 is not seen clearly in the $\DDp$ line shape is as follows:
    The near-equality of the four-point couplings $C_0$ and $C_1$ is obtained, which implies the isospin-independent interaction such as the one obtained by the $\sigma$-exchange force \cite{Sakai:2023syt} to dominate. 
    Moreover, since $C_0$ and $C_1$ are nearly equal, the off-diagonal elements of the potential matrix become almost zero. 
    Consequently, in the particle basis the $\uc$ channel effectively decouples from the $\lc$ channel.
    In addition, the production rates $\beta/\alpha$ is tiny.
    Therefore $\uc$ bound-state pole does not manifest as a clear peak in the $\DDp$ invariant-mass distribution.
    Instead, one would expect this state to produce a distinct signal in channels to which it couples directly, such as $D^0D^+\gamma$ and $D^0D^+\pi^0$ \cite{Ling:2021bir,Chen:2021vhg}. 
    \item[Mol.\,2:] 
    In this scenario, we obtain two bound states, State 1' and State 2', as shown in Fig.\,\ref{fig:spectrum}. 
    By using the same procedure in Mol.1, we obtained State 1' located just 372 keV below the $\lc$ threshold, while this state is realized as the $\uc$ bound state with a binding energy 1.78 MeV.
    State 2' is located 6.16 MeV below the $\lc$ one, associated with the $\lc$ channel just below the $\DDp$ threshold.
    In this scenario, the four-point contact interaction is strengthened compared to the one in the Mol.\,1 parameter set, and the production rates are nearly isospin symmetric.
    Because the fitted $C_0$ and $C_1$ are similar in magnitude as in Mol.\,1, the interaction seems to be dominated by the isospin-independent component across $I=0$ and $I=1$.
    Relative to Mol.\,1, the enhanced attractive four-point interaction, together with a non-negligible difference $C_1 - C_0$, which strengtherns the off-diagonal terms, makes the $\uc$ bound state feed into the $\DDp$ final state; consequently, the $\uc$-channel peak becomes visible in the experimental $\DDp$ spectrum.
    Furthermore, since State\,2' sits below the $\DDp$ threshold, the strong decay into this channel is kinematically closed---hence its peak lies outside the experimental mass range and remains undetected; instead, it may be observable in the radiative mode $\DDg$.
    We stress that this result is not fully consistent with the assumption adopted in the LHCb analysis in terms of its isospin and its realization as a $\ccud$ ground state \cite{LHCb:2021auc}.
\end{enumerate}

The above two molecular scenarios both reproduce the gross features of the $\DDp$ line shape with comparable quality, but they imply rather different underlying dynamics:
These two scenarios differ in which bound state appears as the peak in the $\DDp$ distribution. 
Specifically, in Mol.\,1 scenario the $\lc$ bound state appears as the peak, whereas in Mol.\,2 scenario the $\uc$ bound state appears as the peak.
Nonetheless, their resulting $\chi^2/\mathrm{d.o.f.}$ values differ by less than $0.1$ and two results are statistically indistinguishable even when compared with the fit including a compact component, indicating that neither the pure molecular scenario nor the mixed Mol.+Compact model can be decisively favored given the current experimental uncertainties.
\begin{figure*}[htbp]
  \centering
  \includegraphics[width=0.9\textwidth]{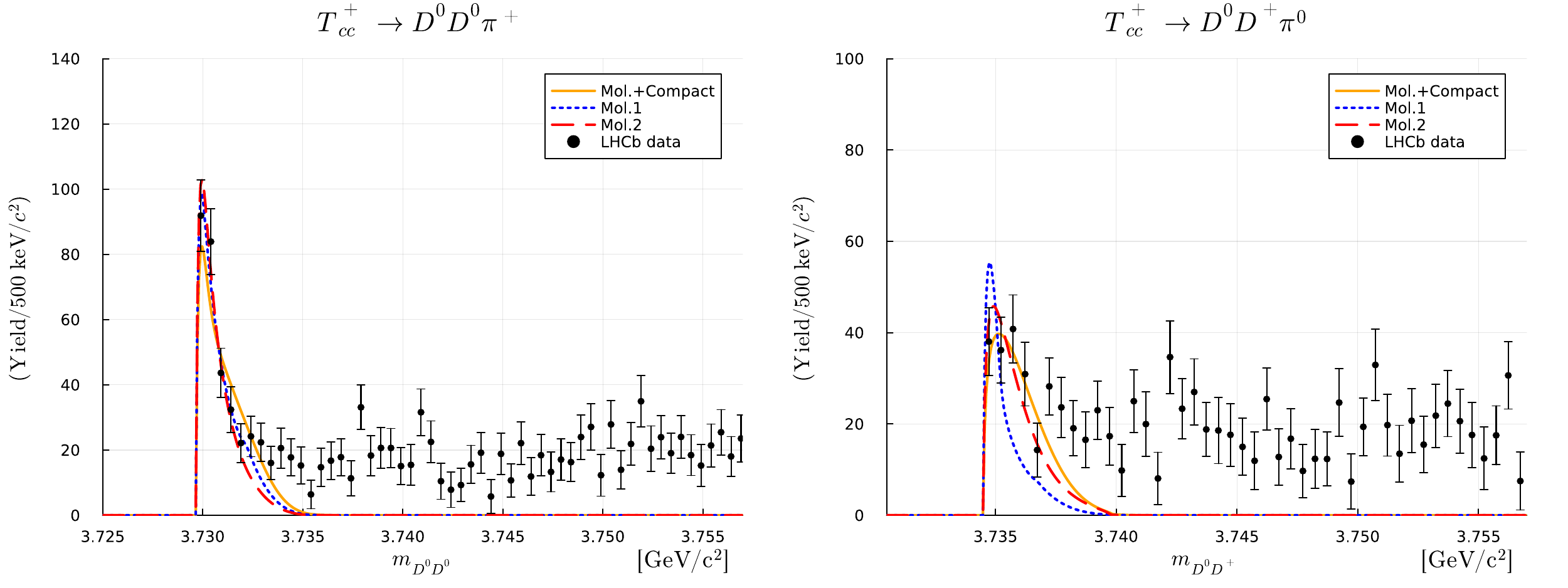}
  \caption{Expected $DD$ invariant-mass distributions derived from the $\DDp$ analysis.
    \textbf{Left:} $T_{cc}^+\!\to \DDp$.
    \textbf{Right:} $T_{cc}^+\!\to D^0D^+\pi^0$ predicted from the $D^0D^+\pi^0$ amplitude computed with the \emph{same} parameter set that was fitted to the 200\,keV-binned $\DDp$ spectrum.
    }
  \label{fig:DD_distribution}
\end{figure*}

Motivated by the above, we examine whether the same best-fitted parameter sets remain consistent with the $DD$ invariant mass distributions.
Fixing all parameters to the best-fitted values in Table\,\ref{tbl:Best Fitted parameters} obtained from the $\DDp$ analysis, we compute the $\DDp$ and $D^0D^+\pi^0$ amplitudes and the corresponding $DD$ line shapes.
Each $DD$ curve is normalized by a $\chi^2$ fit to the first five bins.
Backgrounds at high invariant mass are not modeled, so far from threshold baselines are outside our scope.
For the $D^0D^+$ distribution, the possible $D^0D^+\gamma$ contribution is assumed to be small and is neglected\,\cite{Ling:2021bir,Fleming:2021wmk,Meng:2021jnw}.

In the left panel of Fig.\,\ref{fig:DD_distribution} ($D^0D^0$), all three scenarios reproduce the hallmark near--threshold behavior---a sharp rise followed by a rapid falloff.
The most visible differences just after the peak are governed by the production ratio $\beta/\alpha$ and by channel mixing controlled by $C_1 - C_0$.

In the right panel of Fig.\,\ref{fig:DD_distribution} ($D^0D^+$), the same parameter sets are propagated to the $D^0D^+\pi^0$ amplitude.
Here the scenario dependence is clearer in the peak height and the early--tail damping:
in Mol.\,1 (blue dotted), $C_0 \simeq C_1$ weakens the off--diagonal potential and the $\lc$--dominant pole controls the shape, yielding a comparatively sharper peak and a faster falloff;
in Mol.\,2 (red dashed), a stronger contact attraction together with more symmetric production weights enhances the pole dominated by the $\uc$ channel, producing a somewhat slower initial falloff than in Mol.\,1;
the Mol.+Compact case (orange solid), with a single near--threshold pole, provides a stable description of the $D^0D^+$ peak and its slowly falling tail without over--emphasizing either channel.

Across both panels, given that far from threshold baselines lie outside the model's intended scope, our assessment focuses on the near-threshold region in each panel.
In this window, the most informative features are the local slope at threshold, the peak height relative to nearby data points, and any gentle shoulder in the initial post-peak falloff, all of which are governed by the pole positions and by the strength of channel mixing.
Evaluated on this basis, the three scenarios remain indistinguishable in the current $DD$ line shapes, and none can be excluded.
More detail data near threshold could distinguish these different scenarios and advance our understanding of $T^+_{cc}$.

\textit{Summary}---
We developed the model that considers the mixing of the compact $T_{cc}^+$ and the $DD^*$ molecular states, and analyzed the $\DDp$ lineshape.
We showed three scenarios to explain the peak in the current experimental data corresponding to $\Tcc$:
In the Mol.+Compact scenario, the peak structure is attributed to a compact $T_{cc}^+$ state located near the $DD^*$ threshold.
In addition, we also obtained the pure molecular scenarios Mol.\,1 and Mol.\,2. 
In Mol.\,1, the amplitude develops two bound-state poles.
One is $\lc$ bound state corresponding to the clear peak in the $\DDp$ distribution.
The other is $\uc$ bound state which effectively decouples from the $\DDp$ final state, so its bound-state pole does not emerge as a clear peak in the $\DDp$ distribution.
In Mol.\,2, more attractive four-point contact interaction developed two bound-state poles with larger binding energies.
The $\uc$ bound state appeared as the peak in the experimental $\DDp$ spectrum, while the $\lc$ bound state lies below the $\DDp$ threshold and outside the measured range, potentially visible in the radiative mode $\DDg$.
We would like to stress that, although there is only one pole corresponding to $I = 0$ compact tetraquark state in Mol.+Compact senario, two poles in Mol.1 and Mol.2 are both mixed states of $I = 0$ and $I = 1$.

All three scenarios reproduce the observed line shape with similar quality, and the $\chi^2/\mathrm{d.o.f.}$ values differ by less than 0.1.
Therefore, due to the limited sensitivity of current experimental data, it remains inconclusive whether the observed $T_{cc}^+$ is predominantly of compact, molecular, or mixed nature.

Furthermore, to assess the validity of the three distinct scenarios, we examined the consistency of the two $DD$ spectra. 
From this perspective, all three parameter sets reproduce the characteristic near-threshold behavior of the $DD$ distributions—a sharp turn-on followed by a rapid falloff—and, within current experimental uncertainties, none exhibits a deviation large enough to warrant rejection. 
Consequently, at present the three scenarios remain statistically indistinguishable.
The three scenarios could be distinguished by the increased statistics data.

\textit{Acknowledgments}---
This work was financially supported by JST SPRING, Grant Number JPMJSP2125.
S.A. would like to take this opportunity to thank the “THERS Make New Standards Program for the Next Generation Researchers.”
This work is in part supported by Grants-in-Aid for Scientific Research under Grant Numbers JP20K14478 (Y.Y.), 23H05439, 24K07045 (M.H.), and the RCNP Collaboration Research Network program as the project number COREnet-056 (Y.Y.).

\bibliography{apssamp}% Produces the bibliography via BibTeX.

\end{document}